\documentclass[aps,prd,nofootinbib]{revtex4}
\usepackage[colorlinks=true, pdfstartview=FitV, linkcolor=blue, citecolor=red, urlcolor=magenta]{hyperref}
\usepackage{graphicx}
\usepackage{latexsym}
\usepackage{amsmath}
\usepackage{amsfonts}
\usepackage{amssymb}
\usepackage{caption}
\usepackage{graphics}
\usepackage{color}
\usepackage{epsfig}
\usepackage{graphicx}
\usepackage{subfigure} 
\usepackage{wrapfig} 
\usepackage{epstopdf}
\usepackage{verbatim}

\newcommand{\be}{\begin{equation}}
\newcommand{\ee}{\end{equation}}

\newcommand{\fr}{\frac}
\newcommand{\pa}{\partial}

\DeclareMathOperator{\diag}{diag}

\def\bd{\bar{d}}
\def\td{\tilde{d}}

\def\m{\mu}
\def\n{\nu}
\def\r{\rho}
\def\s{\sigma}
\def\l{\lambda}
\def\g{\gamma}
\def\a{\alpha}
\def\b{\beta}
\def\w{\omega}

\def\t{\tau}
\def\k{\kappa}

\def\pa{\partial}

\def\L{\Lambda}

\def\D{\Delta}

\def\P{\Pi}

\def\la{\label}
\def\cal{\mathcal}

\begin{document}

\title{Consistency analysis of a CPT-even and CPT-odd lorentz-violating effective field theory in the electrodynamics at Planck scale by an influence of a background isotropic field}



\author{$^{1,2}$T. A. M. Sampaio}
\email{thiago.muniz@ifsertao-pe.edu.br}

\author{$^{1}$E. Passos}
\email{passos@df.ufcg.edu.br}


\affiliation{$^{1}$Departamento de F\'{\i}sica, Universidade Federal de Campina Grande,\\
Caixa Postal 10071, 58429-900, Campina Grande, Para\'{\i}ba, Brazil.}
\affiliation{$^{2}$Instituto Federal de Educa\c{c}\~ao, Ci\^encia e Tecnologia do Sert\~ao Pernambucano - campus Salgueiro,\\ Caixa Postal 56000-000, Salgueiro, Pernambuco, Brazil.}


\begin{abstract}
Based on the motivation that some quantum gravity theories predicts the Lorentz Invariance Violation (LIV) around Planck-scale energy levels, this paper proposes a new formalism that addresses the possible effects of LIV in the electrodynamics. This formalism is capable of changing the usual electrodynamics through high derivative arbitrary mass dimension terms that includes a constant background field controlling the intensity of LIV in the models, producing modifications in the dispersion relations in a manner that is similar to the Myers-Pospelov approach. With this framework, it was possible to generate a CPT-even and CPT-odd generalized modifications of the electrodynamics in order to study the stability and causality of these theories considering the isotropic case for the background field. An additional analysis of unitarity at tree level was considered by studying the saturated propagators. After this analysis, we conclude that, while CPT-even modifications always preserves the stability, causality and unitarity in the boundaries of the effective field theory and therefore may be good candidates for field theories with interactions, the CPT-odd one violates causality and unitarity. This feature is a consequence of the vacuum birefringence characteristics that are present in CPT-odd theories for the photon sector.


\end{abstract}
 \maketitle


\section{Introduction}

Einstein's theory of General Relativity (GR) and the standard model of particle physics represents together the most elegant and successful description of nature that exists to date in physics. The first theory describes gravity at the classical level, while the second encompasses all other phenomena involving interactions between particles at the quantum level. It is currently expected that in energies close to the Planck scale, more precisely, the Planck mass $M_P = 1.22 \times 10^{19} GeV$ \footnote{In the natural units system, Planck's energy, and Planck's mass are indistinguishable.} These two field theories are based on a single, unified, quantum-consistent theory of nature \cite{kostelecky2004}.

The current candidates for Quantum Gravity (QG) theory still face enormous difficulties regarding the possibility of direct experimental confirmation, since any experiment aimed at identifying the quantum nature of gravity or space-time would require energies beyond the Planck scale, which is still far from being achieved with current particle accelerators, such as the Large Hadron Collider (LHC). However, there is the possibility of indirect detection of possible effects from QG at low energies \cite{liberati2009}.
 
One of the possible windows to detect these experimental relics would be through the spontaneous breaking of space-time symmetries. In the Planck scale, it is believed that space-time ceases to assume a continuous form to assume a discrete or granular form, with the presence of a minimum fundamental length $l_P = 1.6 \times 10^{-35} m$, called the Planck length. This may suggest that one of the fundamental symmetries of physics, the Lorentz symmetry, can be broken at this energy level. In fact, it is expected that the Lorentz Invariance Violation (LIV) effect, if it exists, is very small in a low-energy regime due to various experimental constraints \cite{liberati2009}, however, it is suggested by Collins \textit{et al} \cite{collins2004} that small LIV effects occurring on the Planck scale can be detected at a low energy level.

In spite of the important role played by Lorentz symmetry in fundamental particle physics, in the last two decades there has been great motivation for studies that consider the possible violation of this symmetry \cite{belich2007}. In fact, the main candidate QG theories, such as loop quantum gravity \cite{ashtekar1986,rovelli1997}, the Horava-Lifshitz gravity \cite{horava2009} and string theory \cite {kostelecky1989} predict that in some cases this may occur in nature.

After Carroll, Field and Jackiw \cite{carroll1990} use a background fourvector \footnote{An external field of unknown origin} to study possible effects of LIV on Maxwell's electrodynamics, Colladay and Kostelecky \cite{colladay1997,colladay1998} elaborated an Effective Field Theory (EFT) in order to investigate LIV in all sectors of the standard model. This formalism became known as the Standard Model Extension (SME).

The mechanism of the EFT with LIV is to add terms with operators that violate the Lorentz invariance in the respective Lagrangian of each model, in order to evaluate possible modifications in the equations of motion and in the dispersion relations of the theory. This method is quite effective for phenomenological purposes, since it provides a sufficiently robust set of rules for describing LIV effects without the need to know details about any fundamental theory of QG \cite{liberati2009}. In fact, many QG models can be reduced to an EFT with LIV, as shown in Refs.\cite{gambini1998,carroll2001,burgess2002}.

The LIV operators may be renormalizable, i.e. with mass dimension 3 or 4 (on which the SME concentrates), or they may be non-renormalizable, having dimensions equal to or greater than 5. The LIV operators of dimension 5 were first introduced in pioneering work by Myers and Pospelov, who discovered high-order modified dispersion relations from an EFT for the scalar, fermionic, and electromagnetic sectors. Non-renormalizable operators have generally gained more prominence in recent years because of the possibility that terms can be suppressed by Planck's energy/mass scale. In this way, these operators could somehow probe possible relics of effects arising from phenomena in the QG scale. In this work, we will propose a formalism to study LIV using a differentiated EFT model through an isotropic background vector.

Several papers have focused on working with modified dispersion relations (which may be from adjacent EFT) to study possible impacts that LIV can induce phenomenologically, as seen in Refs\cite{jacobson2001, shao2010}. The most common example of such concepts in phenomenology is astrophysical or cosmological events with the emission of highly energetic particles. Such events may promote a cumulative LIV effect due to the enormous distances from the sources and the expansion of the Universe, and may bring hope to validate or invalidate LIV theories. Some authors have pointed out that an LIV effect occurring in the production of electromagnetic waves, specially in gamma-ray bursts \cite{amelino2000, castorina2004, shao2011, couturier2013, kislat2017} can result in considerable changes in the photons propagation from distant sources, and may even suggest birefringence effects in the vacuum \cite{carroll1990}.


The outline of this paper is as follows. In Sec. II, we review the famous Myers-Pospelov model for electrodynamics, pointing out some peculiarities of the isotropic case. In Sec. III, we introduce a new formalism for LIV CPT-even and CPT-odd operators with an arbritrary mass dimension in Maxwell's electrodynamics, creating modified wave equations and dispersion relations. We also did an consistency analysis  of the modified theories through stability, causality, and unitarity. In Sec. IV we present the final considerations. In this work, we use the natural unit system, where $c = \hbar = 1 $. The background metric used is the Minkowski metric, with the convention $\eta_{\m\n} = \diag (1, -1, -1, -1) $.

\section{The Myers-Pospelov photon-sector model revisited}

In order to study LIV effects on the Planck scale, Myers and Pospelov elaborated an EFT based on high-order operators with the presence of a constant fourvector interacting with the scalar, fermionic and electromagnetic fields. They proposed the construction of modified lagrangians which can be added in the usual lagrangian of each sector in order to obtain modified dispersion relations. These lagrangians obey the following criteria: (i) they are quadratic in the fields; (ii) are invariant by gauge transformations; (iii) are invariant by Lorentz transformations, except for the presence of an external fourvector $u^\l $; (iv) are not reducible to low dimension operators by the equations of motion; (v) are not reducible to a total derivative; (vi) have one more derivative than the usual term.

The motivation for creating a high derivative EFT that violates Lorentz invariance lies in the fact that the terms can be suppressed by a certain scale of energy where quantum gravitational effects may appear. At this energy level, space-time is expected to take a discrete form, thus justifying that the Lorentz symmetry, being a continuous symmetry, would be broken in this scenario.

The modified Myers-Pospelov term for the photon sector consists of an CPT-odd term, that is, it violates the CPT invariance by a signal. The term proposed by them is given by \cite {myers}
\be \la{lagmp1}
\mathcal{L}_{MP} = \fr{\xi}{2M_P} F_{\m\a} u^\a u_\b (u \cdot \pa ) \tilde{F}^{\m\b} .
\ee
where $\xi$ is a dimensionless parameter that controls the LIV, $u^\a$ is the background fourvector, $M_P$ is the Planck mass and $\tilde{F}^{\m\b} = \fr{1}{2} \epsilon^{\m\b\r\s} F_{\r\s}$ is the dual Maxwell tensor. By adding this lagrangian to the usual Maxwell lagrangian plus the source term one can get the following equation of motion in the Lorentz gauge \cite{reyes2010}:
\be \la{eommyers}
\Box A^\n + 2 \fr{\xi}{M_P} \epsilon^{\n\a\l\s} u_\a (u \cdot \pa )^2 \pa_\l A_\s = j^\n
\ee
which produces, in the absence of sources ($j^\n = 0$), the covariant dispersion relation 
\be \la{rdmp}
k^2 \pm \fr{\xi}{M_P} ( u \cdot k )^2 [(u\cdot k)^2 - u^2 k^2 ]^{1/2} = 0.
\ee
Myers and Pospelov \cite{myers} has considered for analysis the time-like case, which is $u_\l = (1, \vec{0})$. This is also called isotropic model \footnote{because it guarantees that the external field do not provide preferential spatial directions.}. Thus, for $|\vec{k}| < M_P/(2\xi )$, the (\ref{rdmp}) can be approximated by
\be
\w \approx |\vec{k}| - \l \fr{\xi}{M_P} |\vec{k}|^2
\ee
which may induces a sub-luminal group velocity for photons that are under the LIV effect.

\section{High derivative generalized LIV extension for electrodynamics}

In this section, we will create lagrangians with arbitrary mass dimension $d \ge 3$ for the electromagnetic sector. Such terms must comply with all the criteria established by Myers and Pospelov, except (vi), since we want to generate high derivative terms in a generalized way.

The effects of the LIV can be inserted from a modification of the flat space-time metric tensor $\eta_{\a\b}$. This modification can consist of adding a non-dynamic fourvector $u_\l$ to the metric, which we will call external fourvector (or background field). This approach can be given in the form
\be \la{metrica}
\eta_{\a\b} \to \bar{\eta}_{\a\b} = \eta_{\a\b} - \xi u_\a u_\b ,
\ee
where $\xi$ is a dimensionless parameter that controls the intensity of the LIV. By setting the parameter to be positive, the minus sign will ensure that the modified theory does not have superluminal behavior.

Now we are going to analyze the impact of this perturbation on the metric in the usual Maxwell lagrangian without source term, which can be written in terms of the dual tensors of the electromagnetic field in the form
\be \la{lagmaxwell}
\cal{L}_{Maxwell} = \fr{1}{4} \tilde{F}^{\m\n} \tilde{F}^{\r\s} \eta_{\m\r} \eta_{\n\s}.
\ee
Inserting Eq.(\ref{metrica}) in Eq.(\ref{lagmaxwell}), we retrieve the Maxwell lagrangian added with the following extension:
\be \la{lagext}
\cal{L}_{ext} = - \fr{\xi}{2} \tilde{F}^{\m\n} \tilde{F}^{\r\s} \eta_{\m\r} u_\n u_\s .
\ee
In this sense, the effects of LIV are completely characterized by the presence of the $u_\l$ fourvector. Note that the above lagrangian is invariant by gauge transformation $\delta A_\m = \pa_\m \L$. The contribution given in (\ref{lagext}) is known in the literature as aether-like Lorentz violation term \footnote{This term originally appeared in the study of theories with extra dimensions.} \cite{carroll2008}.

By define the operator 
\be \la{oppi}
\Pi_{\m\n} \equiv \epsilon_{\m\n\r\s} u^\r \pa^\s ,
\ee
witch is anti-symmetric and has the properties $\pa_\m \Pi^{\m\n} = 0$ and $u_\m \Pi^{\m\n} = 0$, besides the comutation propertie $\P_{\m\n}\P_{\r\s} = \P_{\r\s}\P_{\m\n}$, we can rewrite the action associated with (\ref{lagext}) in a compact form in terms of the gauge fields up to a surface term as
\be \la{Sext}
S_{ext} = - \int d^4x \fr{\xi}{2} A_\m \P^\m_{\;\;\r} \P^{\r\n} A_\n .
\ee
It is possible to see that this term is invariant by CPT transformation, and thus is classified as CPT-even.

Now we can elaborate a generalization to LIV operators with arbritary mas dimension $d$ in the form
\be
\xi \P^\m_{\;\;\r} \P^{\r\n} \to \fr{\xi_{(d)}}{M_P^{d-4}} \bigg( \P^\m_{\;\;\r} \bigg)^l  \bigg( \P^{\r\n} \bigg)^m.
\ee
where $M_P$ is Planck mass, and $d = 2 + l + m$ is the mass dimension of the operator. The notation $\big( \P_{\m\n} \big)^l$ refers to a contraction of $l$ tensors $\P_{\m\n}$, with the first index of an operator contracted with the second index of the subsequent. Thereby, we can obtain the general action
\be
S_{ext} \to \tilde{S}_{ext} = - \int d^4x  \fr{\xi_{(d)}}{M_P^{d-4}} A_\m  \bigg( \P^\m_{\;\;\r} \bigg)^l  \bigg( \P^{\r\n} \bigg)^m A_\n .
\ee
One can show the following proprietries of $\Pi_{\m\n}$ operator:
\begin{align} \la{pi1}
&\Pi^{\m\n} \P_{\n\m} = 2 \hat{D} ; \nonumber\\
&\Pi_{\m\k} \Pi^{\k\n} = \hat{D} \delta^\n_\m - u^2 \pa^\n \pa_\m + u^\n (u \cdot \pa ) \pa_\m - u^\n u_\m \Box + (u \cdot \pa ) u_\m \pa^\n , \nonumber\\
\end{align}
where $\hat{D} \equiv u^2 \Box - ( u \cdot \pa )^2$.

Using the second relation of Eq.(\ref{pi1}), it is easy to see that a contraction between three operators $\P^{\m\n}$ will result in
\be \la{3pi}
\P^{\m\r} \P_{\r\l} \P^{\l\n} = \P^{\m\r} \hat{D} \delta^\n_\r = \P^{\m\n} \hat{D}.
\ee

According (\ref{pi1}) and (\ref{3pi}), we were able to elaborate table 2.1, that shows the format of the LIV operators and their respectives classifications under CPT transformations. The CPT-odd terms violates CPT invariance by a minus sign. It is interesting to see that, if we perform the transformation $A_\r \to \fr{1}{M_P} \P_{\r\s} A^\s$ in a dimension $d$ term, one obtains a dimension $d+1$ term.

\begingroup \captionof{table}{LIV operators in electrodynamics} \endgroup
\begin{center} 
\begin{tabular}{|c|c|c|}
\hline
Dimension (d) & Operator & CPT \\
\hline
$d=4$ & $\xi_{(4)} A_\m \P^{\m\r} \P_\r^{\;\;\n} A_\n$ & Even \\
\hline
$d=5$ & $\fr{\xi_{(5)}}{M_P} A_\m \P^{\m\n} \hat{D} A_\n$ & Odd \\
\hline
$d=6$ & $\fr{\xi_{(6)}}{M_P^2}  A_\m \P^{\m\r} \P_\r^{\;\;\n} \hat{D} A_\n$ & Even \\
\hline
$d=7$ & $\fr{\xi_{(7)}}{M_P^3}  A_\m \P^{\m\n} \hat{D}^2 A_\n$ & Odd \\
\hline
$d=8$ & $\fr{\xi_{(8)}}{M_P^4}  A_\m \P^{\m\r} \P_\r^{\;\;\n} \hat{D}^2 A_\n$ & Even \\
\hline
\vdots & \vdots & \vdots \\
\hline
\end{tabular} 
\end{center}

From the table, we can see that the operators can be constructed respecting basically a format for CPT-even and another format for CPT-odd, only increasing successively the powers of $\hat{D}$ operator. It is interesting to see that the CPT-odd term can be seen as a high derivative generalization of the CFJ term, which is a term with mass dimension $d=3$, and can be written in our notation as $\xi_{(3)} M_P A_\a \P^{\a\b} A_\b$.

In order to include the CFJ term, we can write the generalized CPT-even and CPT-odd actions for $d \ge 3$ separately in the forms:

\begin{align} \la{lagpargeral}
S^{CPT-even}_{(\bd)} &= - \fr{1}{2} \int d^4 x  \fr{\xi_{(\bd)}}{M_P^{2n}} A_\m \P^{\m\r} \P_\r^{\;\;\n} \hat{D}^n A_\n ;
\\ \la{lagimpargeral}
S^{CPT-odd}_{(\td)} &= - \fr{1}{2} \int d^4 x  \fr{\xi_{(\td)}}{M_P^{2n-1}} A_\m \P^{\m\n} \hat{D}^n A_\n ,
\end{align}
where $n = (0, 1, 2, 3, \cdots )$, $\td = 3+2n$ and $\bd = 4 + 2n$. For $n=0$, the odd term restores the CFJ term, while the even part restores the dimension $d=4$ term (\ref{Sext}). For $n > 0$, we obtain the high derivative terms in which we are interested. The lagrangians densities contained in these actions satisfy the five Myers-Pospelov criteria previously established.

We can rewrite (\ref{lagpargeral}) and (\ref{lagimpargeral}) in terms of electromagnetic dual tensor
\begin{align}
S_{(\bd)}^{CPT-even} &= \int d^4 x \frac{\xi_{(\bd)}}{2 M_P^{2n}} u^\a u_\b \tilde{F}_{\m\a} \hat{D}^n \tilde{F}^{\m\b} ; \\
S_{(\td)}^{CPT-odd} & = \int d^4 x \fr{\xi_{(\td)}}{2 M_P^{2n-1}} u_\s A_\m \hat{D}^n \tilde{F}^{\m\s}
\end{align}

The operators included in the above actions can be written as pieces of CPT-even $(K_F)$ and CPT-odd $(K_{AF})$ operators present in SME framework, that was created by Ref.\cite{colladay1998}, and generalized by Refs.\cite{kosteleckymewes2007,kosteleckymewes2009}. The relation between these frameworks was exposed in Appendix A. 
Therefore, the framework proposed in this work, which started from the motivation and the criteria originally used by Myers and Pospelov, consists of a very useful simplification of the generalized SME for the photon sector, since in this way, we can analyze the entire behavior of the LIV through the nature of the background field.

In this work, we will focus on investigating Lorentz violations by boost transformations. For this, we will always assume that the background fourvector $u^\l$ is time-like, having the form $u^\l = (1, \vec{0})$, with the time component normalized in terms of the $\xi$ parameter. With this approach, we are excluding anisotropic effects that may exist in space-time due to LIV.


\subsection{CPT-even extension}

In this section we are going to analyse the modified electrodynamics associated with CPT-even term (\ref{lagpargeral}). In this way, we start from the action
\be \la{acpar}
S = \int d^4 x \bigg( - \fr{1}{4} F_{\m\n} F^{\m\n} + j_\m A^\m + \fr{\xi_{(\bd)}}{2M_P^{2n}} A_\m \P^{\m\r} \P_\r^{\;\;\n} \hat{D}^n A_\n  \bigg).
\ee
The equation of motion that can be obtained is
\begin{align} \la{eompar1}
\fr{\delta S}{\delta A^\m} &= \Box A_\m - \pa_\m \pa^\n A_\n  - \frac{\xi_{(\bd)}}{M_P^{2n}} \Pi_{\m\l} \Pi^{\l\a}\hat{D}^n A_\a + j_\m \nonumber\\
&= \Box A_\m - \pa_\m \pa_\n A_\n  + \frac{\xi_{(\bd)}}{M_P^{2n}} \hat{D}^n \bigg( - \hat{D} A_\mu + u_\n u^\mu \Box A^\n - u^\mu \partial_\nu (u\cdot \partial ) A^\nu - u_\nu \partial^\mu (u \cdot \partial ) A^\nu \nonumber\\
& + u^2 \partial^\mu \partial_\nu A^\nu \bigg) + j_\m = 0.
\end{align}
To investigate the propagation of electromagnetic waves and obtain the covariant dispersion relation of the theory, we can consider a space in the absence of sources, that is, $j_\m = 0 $. Using the Lorentz gauge $\partial_\m A^\m = 0$ and the axial gauge $u_\m A^\m = 0$ on (\ref{eompar1}), we get
\be \la{eompar2}
\bigg( \Box - \frac{\xi_{(\bd)}}{M_P^{2n}} \hat{D}^{n+1} \bigg) A_\m = 0.
\ee 
In the momentum space ($\partial_\mu \to - i k_\mu$ ), we get the following covariant dispersion relation
\be \la{rdpar1}
k^2 + \frac{\xi_{(\bd)}}{M_P^{2n}} D^{n+1} = 0,
\ee
in which $D = \hat{D}(\partial_\mu \to - i k_\mu ) = (u \cdot k )^2 - u^2 k^2$.

In the isotropic model, where $u^\a = (1, \vec{0} )$, the equation (\ref{rdpar1}) results in
\be \la{rdpar2}
\w = \pm  |\vec{k}| \sqrt{ 1 - \frac{\xi_{(\bd)}}{M_P^{2n}} |\vec{k}|^{2n}}.
\ee
It is important to note that usual case $\w = \pm |\vec{k}|$ is recovered when $\xi_{(d)} \to 0$. The group velocity obtained from (\ref{rdpar2}) is
\be \la{vgpar}
v_{g} = \frac{d\w}{d|\vec{k}|} =  \frac{1 - \frac{\xi_{(\bd)}}{M_P^{2n}} (n+1) |\vec{k}|^{2n}}{\sqrt{ 1 - \fr{\xi}{M_P^{2n}} |\vec{k}|^{2n}}},
\ee
while the phase velocity is expressed by
\be \la{vfpar}
v_{p} = \fr{\w}{|\vec{k}|} = \sqrt{1 - \fr{\xi_{(\bd)}}{M_P^{2n}} |\vec{k}|^{2n}},
\ee 
thus characterizing a dispersive medium.

Now, we will analyze the consistency of the theory in the isotropic model model through stability, causality and unitarity. These three conditions are indispensable so that the theory can be quantized in accordance with the methods of quantum field theory.

\subsubsection{Stability}

The stability is guaranteed when the energy of the particle is real and positive. This occurs if $\w^2 \ge 0$. Starting from the dispersion relation of CPT-even modified electrodynamics (\ref{rdpar2}), we see that the theory is stable if $\xi_{(\bd)} \le \fr{M_P^{2n}}{|\vec{k}|^{2n}}$. Then, the CPT-even preserves stability for any positive values of $\xi_{(\bd)}$ in the interval $0 \le \xi_{(\bd)} \le \fr{M_P^{2n}}{|\vec{k}|^{2n}}$.

For the CPT-even $\bd = 4$ case  ($n=0$), the stability is guaranteed if $\xi_{(\bd=4)} \le 1$. This is largely expected in phenomenology, since in this case the LIV effects are expected to be very small and fully controlled by the parameter $\xi$. For high-derivative CPT-even cases with $\bd \ge 6$ ($n \ge 1$), if we assume that LIV effects occurs exactly at the Planck scale, then we can consider $\xi_{(\bd \ge 6)}= 1$, and the stability condition will be $|\vec{k}| \le M_P$. In fact, the phenomenology requires that the momentum $|\vec{k}|$ of the particles be smaller than the Planck mass $M_P$. Thereby, the isotropic CPT-even modification in the boundaries of effective field theory is stable.

\subsubsection{Causality}

Causality is maintained when there are no superluminal (tachionic) modes in the theory, that is, when the conditions $v_g \le 1$ and $v_{front} \le 1$ are satisfied, in which $v_{front}$ is called front velocity, and it is defined by the maximum phase velocity over all possible values of $|\vec{k}|$.

Starting from the (\ref{vgpar}), we see that the subluminal group velocities are guaranteed if the following inequation was satisfied:
\be
1 - \fr{\xi_{(\bd)}}{M_P^{2n}} (n+1) |\vec{k}|^{2n} \le \sqrt{ 1 - \fr{\xi_{(\bd)}}{M_P^{2n}} |\vec{k}|^{2n}},
\ee
what happens for $0 \le \xi_{(\bd)} \le \fr{M_P^{2n}}{|\vec{k}|^{2n}}$, which agrees with the stability condition.

Using the phase velocity (\ref{vfpar}), for $|\vec{k}| \sim M_P$, the front velocity becomes
\be
v_{front} = \sqrt{1 - \xi_{(\bd)}}
\ee
Therefore, causality is generally maintained for $0 \le \xi_{(\bd)} \le 1$. In general this interval for $\xi$ parameter is strongly supported by what is expected of phenomenology, as we discussed above in the stability regime.

We can consider (\ref{rdpar2}) taken until the first order of approximation in $|\vec{k}|$,
\be \la{approx1}
\w \approx \pm |\vec{k}| \bigg( 1 - \fr{1}{2} (g |\vec{k}|)^{2n} \bigg)
\ee
where\footnote{We had considered an arbritary value of $\xi$.} $g \equiv \xi /M_P$, with $|\vec{k}| < g^{-1}$. By the equation (\ref{approx1}), we can plot a $g\w$ \textit{versus} $g|\vec{k}|$ graph, to the cases $d=4$, $d=6$ and $d=8$, as shown in figure 1.

\begin{figure}[htb]
\begingroup \captionof{figure}{Graph from modified dispersion relations in the CPT-even electrodynamics} \endgroup
\begin{center}
\includegraphics[scale=0.5]{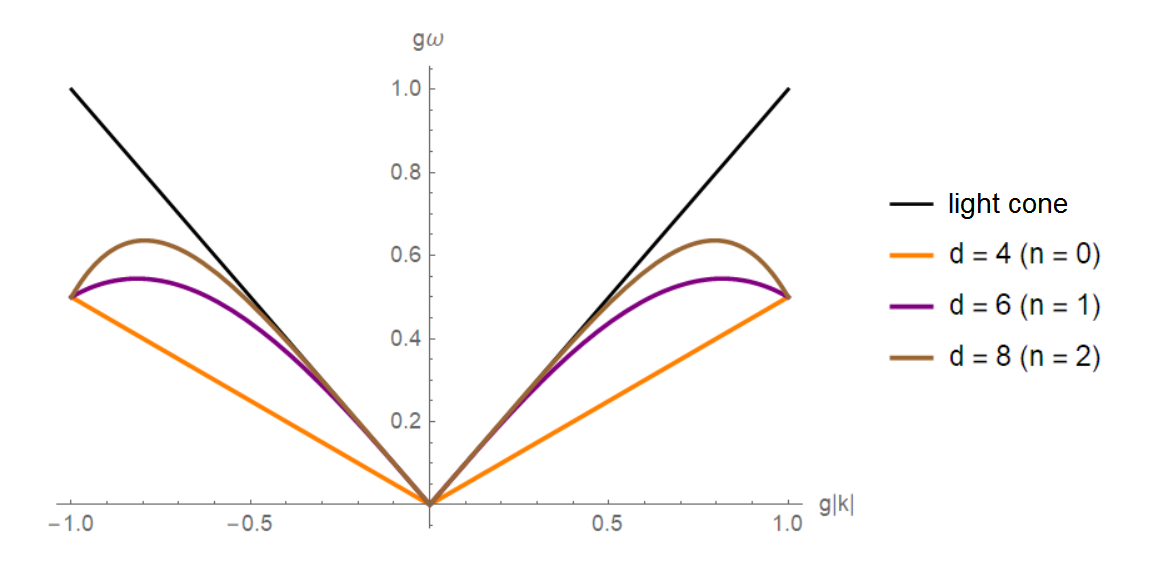}
\label{fig2.1}
\end{center}
\end{figure}

We see that all cases concentrate outside the light cone, which guarantees causality by the sub luminal behavior of the photons. Stability is only guaranteed by small values ​​of momentum $|\vec{k}|$, as previously discussed, since the prolongation of the $d \ge 6$ curves would induce negative energies.
 
\subsubsection{Unitarity}

Unitarity is related to the property of the scattering matrix of quantum field theory. This condition guarantees hermiticity, that is, it ensures that the states defined in the Hilbert space have positive norms. When there are states of negative norm, the theory will be non-unitary, and we have the so-called ghost modes.

The unitarity can be verified through Feynman's saturated propagator of the theory. If its residue is always positive, the theory will be unitary; if it is null, the theory is non-physical, since that does not contribute to computation in the scattering matrix in interactions between particles;  and if it assume negative values, the theory is non-unitary and have ghost modes.

First, let us compute the Feynman propagator associated with the CPT-even theory. For this, we will add  Feynman's gauge fixing term $-\fr{1}{2} (\pa_\m A^\m )^2$ in the action (\ref{acpar}) without the source term. Then, we start from action
\be
S = \int d^4x \bigg( - \fr{1}{4} F_{\m\n} F^{\m\n} - \fr{\xi_{(\bd)}}{M_P^{2n}} A^\s \P_{\s\m} \P^{\m\l} \hat{D}^n A_\l - \fr{1}{2} (\pa_\m A^\m )^2 \bigg) ,
\ee
that results in
\be
S = \int d^4x \fr{1}{2} A^\s \bigg( \Box \delta^\l_\s - \fr{\xi_{(\bd)}}{M_P^{2n}} \P_{\s\m} \P^{\m\l} \hat{D}^n \bigg) A_\l .
\ee
We can rewrite the above action as
\be
S = \int d^4x \fr{1}{2} A_\s \cal{O}^{\l\s} A_\l ,
\ee
where
\be \la{opondapar}
\cal{O}^{\l\s} = \Box \eta^{\l\s} - \fr{\xi_{(\bd)}}{M_P^{2n}} \P^\s_{\;\;\m} \P^{\m\l} \hat{D}^n ,
\ee
is called wave operator. The propagator is the inverse of this operator, and can be obtained by the identity
\be \la{ident}
\cal{O}^{\l\s} \D_{\l\r} = i \delta^\s_\r .
\ee
After some algebra, the propagator obtained in the momentum space is
\begin{align} \la{proppar}
\D_{\l\r} (k) &= \fr{i}{k^2 \bigg( k^2 + \fr{\xi_{(\bd)}}{M_P^{2n}} D^{n+1} \bigg)} \bigg[ - \fr{\xi_{(\bd)}}{M_P^{2n}} D^n [ D + (u \cdot k)^2 ] \fr{k_\l k_\r}{k^2} - k^2 \eta_{\l\r} \nonumber\\ & - \fr{\xi_{(\bd)}}{M_P^{2n}} D^n (u \cdot k) ( u_\l k_\r + u_\r k_\l ) 
- \fr{\xi_{(\bd)}}{M_P^{2n}} k^2 D^n u_\l u_\r \bigg] .
\end{align}
Analysing the denominator, we have two poles: $k^2 = 0$ and $k^2 + \fr{\xi_{(\bd)}}{M_P^{2n}} D^{n+1} = 0$. The first one consists in the usual dispersion relation, while the second one, in the modified dispersion relation found in (\ref{rdpar1}). It is interesting to note that, in the limit $\xi_{(\bd)} \to 0$, we recover the usual photon propagator in Feyman gauge \cite{greiner},
\be \la{propusual}
\D_{\l\r}^{usual}(k) = - \fr{i}{k^2} \eta_{\l\r} .
\ee

Now, the saturated propagator is defined by
\be \la{PS}
SP = J^\l \D_{\l\r} (k) J^\r ,
\ee
where $J^\m$ is a conserved current, that is, $\pa_\m J^\m = 0$, which in the momentum space results in $k_\m J^\m = 0$. Inserting (\ref{proppar}) in (\ref{PS}), we are taken to
\be \la{ps1}
SP = - \frac{ i [ J^2 + \frac{\xi_{(\bd)}}{M_P^{2n}} D^n (u \cdot J)^2 ]}{ k^2 + \frac{\xi_{(\bd)}}{M_P^{2n}} D^{n+1}},
\ee
where $J^2 \equiv J_\m J^\m$. In the isotropic case, the saturated propagator will be
\be
SP = - i \fr{ J^2 + \fr{\xi_{(\td)}}{M_P^{2n-1}} |\vec{k}|^{2n} J_0^2 }{ k^2 + \fr{\xi_{(\td)}}{M_P^{2n-1}} |\vec{k}|^{2n+2} } .
\ee
To simplify the problem we can make the choice $k_\m = (\w , 0 , 0 , k_3 )$ without loss of generality\cite{scatena2014}. With this, we can obtain $J_\m = (J_0 , J_1 , J_2 , \fr{\w J_0}{k_3} )$. Then, the (\ref{ps1}) turns in
\be
SP = \fr{ - i \bigg( J_0^2 - J_1^2 J_2^2 - \fr{\w^2 J_0^2}{k_3^2} + \frac{\xi_{(\bd)}}{M_P^{2n}} k_3^{2n} \bigg) }{ \w^2 k_3^2 ( 1 - \frac{\xi_{(\bd)}}{M_P^{2n}} k_3^{2n} ) } .
\ee
With this, we can computate the residue in the pole $\w^2 \to k_3^2 (1 - \frac{\xi_{(\bd)}}{M_P^{2n}} k_3^{2n} ) $, which leads us to the result
\be
Res[SP]|_{\w^2 \to k_3^2 (1 - \frac{\xi_{(\bd)}}{M_P^{2n}} k_3^{2n} )} = i (J_1^2 + J_2^2 ) ,
\ee
which is always positive. Therefore, the CPT-even modified Maxwell's theory is unitary.


\subsection{CPT-odd extension}

In this section we will study the case of electrodynamics modified by CPT-odd LIV term (\ref{lagimpargeral}). We start from the action
\be \la{acimpar}
S = \int d^4x \bigg( - \fr{1}{4} F_{\m\n} F^{\m\n} + j_\m A^\m - \fr{\xi_{(\td)}}{2 M_P^{2n-1}} A_\m \P^{\m\n} \hat{D}^n A_\n  \bigg) ,
\ee
whose equation of motion is
\be \la{eomimpar1}
\fr{\delta S}{\delta A^\m} = \Box A_\m - \pa_\m \pa_\n A^\n + j_\m - \fr{\xi_{(\td)}}{M_P^{2n-1}} \P_{\m\n} \hat{D}^n A^\n  = 0 .
\ee
Now, considering the vacumm in abscence of sources and the choice of Lorentz gauge $\pa_\l A^\l = 0$,  the (\ref{eomimpar1}) turns in
\be \la{eomimpar2}
\bigg[ \eta_{\m\n} \Box - \frac{\xi_{(\td)}}{ M_P^{2n-1}} \Pi_{\m\n} \hat{D}^n \bigg] A^\n = 0.
\ee
In the momentum space, we obtain
\be \la{eomimpar2}
\bigg[ k^2 \eta_{\m\n} + \frac{\xi_{(\td)}}{M_P^{2n-1}} \tilde{\Pi}_{\m\n} D^n \bigg] A^\n = 0,
\ee
where $\tilde{\P}_{\m\n} = \P_{\m\n}(\pa_\a \to -i k_\a ) = - i \epsilon_{\r\s\m\n} u^\r \pa^\s$. To obtain the covariant dispersion relation, we can multiply the above equation by its conjugate $k^2 \eta^{\r\m} - \frac{\xi_{(\td)}}{M_P^{2n-1}} \tilde{\Pi}^{\r\m} D^n
$. After some algebra, we come to
\be \la{eomimpar4}
\bigg[ (k^2)^2 - \bigg( \fr{\xi_{(\td)}}{M_P^{2n-1}}\bigg)^2 D^{2n+1} \bigg] A^\r = 0.
\ee
and lastly, the dispersion relation
\be \la{rdimpar1}
k^2 - \l \fr{\xi_{(\td)}}{M_P^{2n-1}} D^{n+\fr{1}{2}} = 0,
\ee
in which $\l = \pm 1$ are the two polarization modes of electromagnetic waves in presence of CPT-odd LIV extension.

Again, considering the isotropic case $u_\m = (1, \vec{0})$, (\ref{rdimpar1}) yields
\be \la{rdimpar2}
\w_\pm = |\vec{k}| \sqrt{ 1 + \l \fr{\xi_{(\td)}}{M_P^{2n-1}} |\vec{k}|^{2n-1}}.
\ee
thereby, we can see that the group velocity has the form
\be \la{vgimpar}
v_g = \fr{d\w}{d|\vec{k}|} = \fr{1+ (n+\fr{1}{2}) \l \fr{\xi_{(\td)}}{M_P^{2n-1}} |\vec{k}|^{2n-1}}{\sqrt{1+ \l \fr{\xi_{(d)}}{M_P^{2n-1}} |\vec{k}|^{2n-1}}},
\ee
while the phase velocity is
\be \la{vfimpar}
v_p = \fr{\w}{|\vec{k}|} = \sqrt{1+ \l \fr{\xi_{(\td)}}{M_P^{2n-1}} |\vec{k}|^{2n-1}}.
\ee

it can be seen that the CPT-odd LIV modification imposes two different polarization modes, since the group and phase velocities of the two polarization modes will be different. This suggests that the vacuum itself would behave as a birefringent medium, in addition to being dispersive, as in the case of CPT-even modification. This behavior was already expected, since that our CPT-odd framework consists in a generalization of CFJ and Myers-Pospelov CPT-odd modifications.

\subsubsection{Stability}

From the dispersion relation (\ref{rdimpar2}), the stability condition is $1 + \lambda \xi_{(\td)} \frac{M_P^{2n-1}}{|\vec{k}|^{2n-1}} \ge 0$, that depends of the two polarization modes $\lambda = \pm 1$. For $\lambda = +1$, the stability is guaranteed in all cases, for any positive values of $\xi_{(\td)}$. For $\lambda = -1$, the stability could be maintained for the same reason that we discussed in the CPT-even case. Thereby, the CPT-odd theory is in general stable.

\subsubsection{Causality}

From (\ref{vgimpar}), we see that the subluminal group velocities will be associated to the following inequality
\be
1 + \bigg( n+\fr{1}{2} \bigg) \l \fr{\xi_{(\td)}}{M_P^{2n-1}} |\vec{k}|^{2n-1} \le \sqrt{ 1 + \l \fr{\xi_{(\td)}}{M_P^{2n-1}}                                                                                                                                      |\vec{k}|^{2n-1}} .
\ee
For $\l = -1$, the causality is assured for $0 \le \xi_{(\td)} \le \fr{M_P^{2n-1}}{|\vec{k}|^{2n-1}}$, with $n > 0$, analogously to the CPT-even case. However, for $\l = +1$, the above inequality is not obeyed for any positive values of $\xi_{(\td)}$.

From the phase velocity (\ref{vfimpar}), for $|\vec{k}| \sim M_P$, we obtain
\be
v_{front} = \sqrt{ 1 + \l \xi_{(\td)}},
\ee
what implies that, while $\l = -1$ brings the same causality condition analysed in the CPT-even case, $\l = +1$ however corresponds to a superluminal signal, and therefore a non-causal mode of propagation.

We can consider the analysis of the first-order approximation in $|\vec{k}|$ of the equation (\ref{rdimpar2}),
\be
\w_{\pm} \approx |\vec{k}| \bigg( 1 + \fr{1}{2} \l (y |\vec{k}|)^{2n+1} \bigg),
\ee
where $y \equiv \xi /M_P$, in which $|\vec{k}| < y^{-1}$. We can analyze the causal behavior of the theory by a $y\w$ \textit{versus} $y|\vec{k}|$ graph, for the cases $d=3$, $d=5$ and $d=7$, as shown in Figure 2.

\begin{figure}[htb]
\begingroup \captionof{figure}{Graph of the CPT-odd modified disperion relations.} \endgroup
\begin{center}
\includegraphics[scale=0.5]{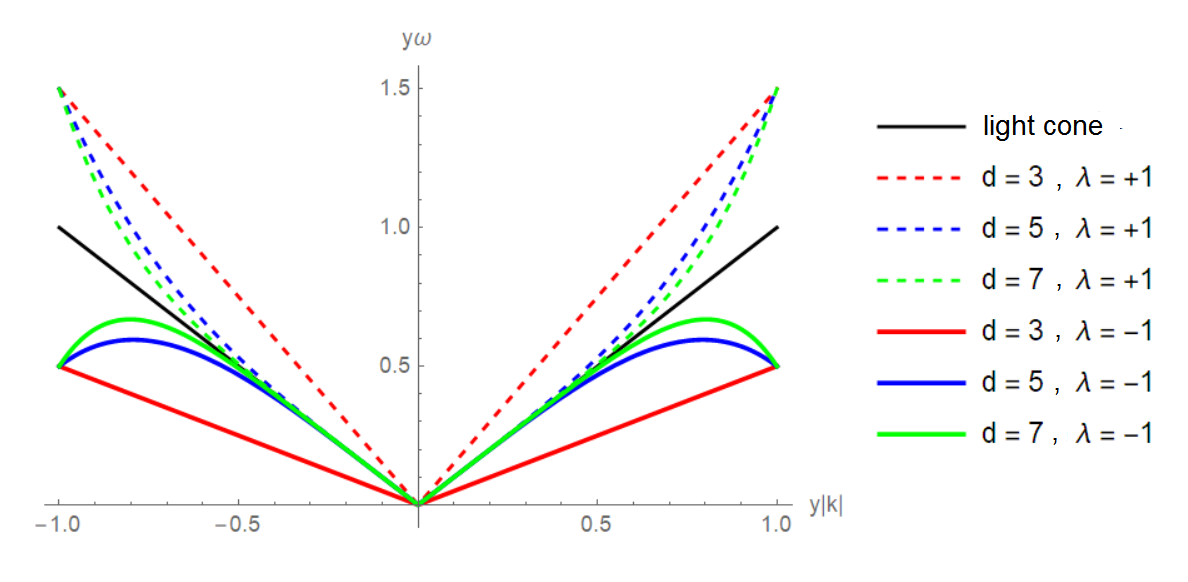}
\label{fig2.2}
\end{center}
\end{figure}

We see that the modes $\l = +1$ are inside the light cone, which indicates a super luminal behavior. Therefore, the modified CPT-odd theory in the isotropic case is non-causal, since one of the polarization modes always induces the existence of tachyons. This analysis agrees with the analysis made by the Ref.\cite{reyes2010}, which shows that the electrodynamics modified with the CPT-odd Myers-Pospelov term is also generally non-causal in the isotropic case.

\subsubsection{Unitarity}

To get the propagator of CPT-odd electrodynamics extension propagator, we consider the action (\ref{acimpar}) without the source term in addition to gauge fixing term
\be
S = \int d^4x \bigg( - \fr{1}{4} F_{\m\n} F^{\m\n} - \fr{\xi_{(\td)}}{M_P^{2n-1}} A_\m\s \P_{\m\n} \hat{D}^n A_\n  - \fr{1}{2} (\pa_\m A^\m )^2 \bigg) ,
\ee
that results in
\be
S = \int d^4x \fr{1}{2} A^\s \bigg( \Box \delta^\l_\s - \fr{\xi_{(\td)}}{M_P^{2n-1}}\P^{\l\s} \hat{D}^n \bigg) A_\l .
\ee
Rewriting the above action as
\be
S = \int d^4x \fr{1}{2} A_\s \cal{\tilde{O}}^{\l\s} A_\l ,
\ee
where
\be \la{opondaimpar}
\cal{\tilde{O}}^{\l\s} = \Box \eta^{\l\s} - \fr{\xi_{(\td)}}{M_P^{2n-1}} \P^{\l\s} \hat{D}^n
\ee
is the wave operator, we can obtain the propagator according to the identity (\ref{ident}). After some algebra, the propagator in the momentum space is
\begin{align} \la{propimpar}
\tilde{\D}_{\l\r} (k) &= - \fr{i}{k^2 \bigg( k^4 - \bigg( \fr{\xi_{(\td)}}{M_P^{2n-1}} \bigg)^2 D^{2n+1} \bigg)} \bigg[ \bigg(\fr{\xi_{(\td)}}{M_P^{2n-1}} \bigg)^2 D^{2n} u^2 k^2 \fr{k_\l k_\r}{k^2} + k^4 \eta_{\l\r} - \fr{\xi_{(\td)}}{M_P^{2n-1}} D^n k^2 \tilde{\P}_{\l\r} \nonumber\\ &- i \bigg( \fr{\xi_{(\td)}}{M_P^{2n-1}} \bigg)^2 D^{2n} (u \cdot k) \bigg( u_\l k_\r + u_\r k_\l \bigg) + \bigg( \fr{\xi_{(\td)}}{M_P^{2n-1}} \bigg)^2 D^{2n} k^2 u_\l u_\r \bigg] .
\end{align}
Analysing that, we have two poles: $k^2 = 0$, which is the usual disperion relation, and $k^4 - \fr{\xi_{(\td)}}{M_P^{2n-1}} D^{2n+1} = 0$, which is the modified dispersion relation (\ref{rdimpar1}). We see that, in the limit $\xi_{(\td)} \to 0$, we retrieve the usual photon propagator (\ref{propusual}). 

The saturated propagator obtained from (\ref{propimpar}) will be
\be
SP = - \fr{ i \bigg[ k^2 J^2 + \bigg( \fr{\xi_{(\td)}}{M_P^{2n-1}} \bigg)^2 |\vec{k}^{4n}| J_0^2 \bigg] }{ (k^2)^2 - \bigg(  \fr{\xi_{(\td)}}{M_P^{2n-1}} \bigg)^2 |\vec{k}|^{4n+2} } .
\ee
Making the choice $k_\m = (\w , 0 , 0 , k_3 )$, we obtain
\be \la{sp2}
SP = \fr{ - i \bigg[ \bigg(\w^2 - k_3^2 \bigg) \bigg( J_0^2 - J_1^2 - J_2^2 - \fr{\w^2 J_0^2}{k_3^2} \bigg) +  \bigg( \fr{\xi_{(\td)}}{M_P^{2n-1}} \bigg)^2 k_3^{4n} J_0^2 \bigg] }{ \w^2 - k_3^2 ( 1 \pm \fr{\xi_{(\td)}}{M_P^{2n-1}} k_3^{2n-1} )} .
\ee
thereby, we have two simple pole $\w \to m^2_\pm $, where $m^2_\pm = k_3^2 ( 1 \pm \fr{\xi_{(\td)}}{M_P^{2n-1}} k_3^{2n-1} )$. The residue from (\ref{sp2}) is
\be
Res[SP]|_{\w^2 \to m^2_\pm} =  \pm \fr{i}{2} ( - J_1^2 - J_2^2 ) .
\ee
Clearly we see that the residue is not positive-definite. Therefore, the theory is not unitary. In this case, the birefringent polarization modes induces the appearance of ghost modes in the model. This result agrees with the  
Ref.\cite{scatena2014}, which showed that the modified electrodynamics with CPT-odd Myers-Pospelov isotropic model also violates the unitarity due to the birefringent modes.

\section{Conclusions}

In this work, we elaborate a new framework in the form of an effective field theory with the creation of modified lagrangians with high derivative operators in order to include the LIV effects through the presence of an isotropic background field in Maxwell's electrodynamics. 

These modified LIV formalism for electrodynamics simplify the consistency analysis of theses theories, in comarision to the SME framework. The dispersion relations in the isotropic case of these models shows that CPT-even modification preserves stability, causality and unitarity in the boundaries of effective field theory. In this way, the CPT-even electrodynamics in isotropic case may be a good candidate for quantum field theories with interactions, and therefore deserve further study in this regard.

However, the CPT-odd electrodynamics violates causality and unitarity, inducting the appearance of ghost and tachionic modes. This feature is probably a consequence of the different polarization modes that appears in the dispersion relation, suggesting that the vacuum has birefringence characteristics. This strange, but interesting possible nature for the vacuum is always present in the discussions about CPT-odd theories for the photon sector, but there is no experimental confirmation for this idea so far.

The phenomenological studies already carried out, mainly with astrophysical experiments of gamma-ray bursts, imposed large restrictions in the values of the parameters $\xi_{d=5}$ and $\xi_{d=6}$, as seen for example in Ref.\cite{abdo2009}, indicating that perhaps the energy scale where the LIV effects may occur are even larger than the scale defined by the Planck mass. This may indicate that Lorentz's invariance may be an exact symmetry of nature, or that the framework of effective field theory may be insufficient to probe such effects \cite{liberati2009}. 

However, it is encouraging that future advances in experimental accuracies of measurements can give us a better perspective on the study of these possible relics of a more fundamental theory of quantum gravity. In a general way, a more in-depth theoretical study in this sense can help us to understand the real origins of this possible space-time symmetry breaking.

{\acknowledgments} We would like to thank to to CNPq and CAPES for partial financial support.

\vspace{1cm}
\appendix
\section{High-derivative operators and the non-minimal Standard Model Extension}

We can verify that Eq.(\ref{lagpargeral}) could be rewritten as a piece of the CPT-even photon term of the SME stablished in Ref.\cite{colladay1998}. In this framework, the CPT-even piece is given by
\begin{align} \la{EMPpar}
\cal{L}^{CPT-even}_{SME} &= -\fr{1}{4} (K_F)_{\m\n\a\b} F^{\m\n} F^{\a\b} \nonumber\\
&=  -\fr{1}{4} (K_F)_{\m\n\a\b} A^\m \pa^\n \pa^\a A^\b.
\end{align}
where the tensor $(K_F)_{\m\n\a\b}$ is antisymmetric in changing the first two and the last two indexes, and symmetric by changing the first pair with the second pair of indexes, that is
\be \la{tensk1}
(K_F)_{\m\n\a\b} = - (K_F)_{\n\m\a\b} = - (K_F)_{\m\n\b\a} = (K_F)_{\a\b\m\n},
\ee
and it has null double trace: $(K_F)^{\r\s}_{\;\;\;\;\r\s} = 0$.

This tensor has 19 independent components, in which 10 are associated to birefringent effects. An interesting way to parametrize the 9 non-birefringent components of $(K_F)$ consists to rewrite it in terms of a symmetric tensor $C_{\m\n}$ in the form \cite{kostelecky1989,hernaski2014,Casana2010a,Casana2010b}
\be \la{tensk2}
(K_F)_{\m\n\a\b} = \fr{1}{2} \bigg( \eta_{\m\a} C_{\n\b} - \eta_{\n\a} C_{\m\b} + \eta_{\n\b} C_{\m\a} - \eta_{\m\b} C_{\n\a} \bigg),
\ee
where
\be \la{tensc}
C_{\t\k} = u_\t u_\k - \fr{1}{2} u^2 \eta_{\t\k}.
\ee
Inserting above equations in the lagrangian given in Eq.(\ref{EMPpar}), we can compare with Eq.(\ref{lagpargeral}) for $d=4$, and infer that $\cal{L}^{CPT-even}_{SME} = - \cal{L}^{CPT-even}_{(4)}$. Based on this, we can write the coefficient $K_F$ in terms of the background fourvector as
\be \la{tensk3}
(K_F)_{\m\n\a\b} = - \xi_{(4)} \epsilon_{\m\n\r\l} \epsilon_{\a\b\s\t} \eta^{\t\l} u^\r u^\s.
\ee

The CPT-odd SME piece for photon sector is given by
\begin{align} \la{EMPimpar}
\cal{L}_{SME}^{CPT-odd} &= \fr{1}{2} \epsilon^{\k\l\m\n} A_\l (K_{AF})_\k F_{\m\n} \nonumber\\
&= \epsilon^{\k\l\m\n} A_\l (K_{AF})_\k \pa_\m A_\n ,
\end{align}
where $(K_{AF})_\k$ is a fourvector with mass dimension. By comparison with Eq.(\ref{lagimpargeral}) for $d=3$, we can rewrite the Eq.(\ref{EMPimpar}) in the form
\be \la{tenskaf}
(K_{AF})_\k = \xi_{(3)} M_\g u_\k .
\ee

The papers in Refs.\cite{kosteleckymewes2007,kosteleckymewes2009} had constructed a generalization in order to include high-derivative operators with mass dimension $d > 4$. The even and odd pieces of this non-minimal SME are written as
\begin{align} \la{EMPpargeral}
\cal{L}_{(\bd)}^{SME-even} = -\fr{1}{4} (\cal{K}^{(\bd)}_F)_{\m\n\a\b} A^\m \pa^\n \pa^\a A^\b ; \\ \la{EMPimpargeral}
\cal{L}_{(\td)}^{SME-odd} = \epsilon^{\k\l\m\n} A_\l (\cal{K}^{(\td)}_{AF})_\k \pa_\m A_\n ,
\end{align}
where
\begin{align} \la{tensktotal1}
(\cal{K}^{(\bd)}_F)_{\m\n\a\b} &=  (K^{(\bd)}_F)_{\m\n\a\b{\r_1 \cdots \r_{(\bd -4)}}} \pa^{\r_1} \cdots \pa^{\r_{(\bd -4)}}, \nonumber\\
(\cal{K}^{(\td)}_{AF})_\k &= (K_{AF}^{(\td)})_\k^{\;\; \r_1 \cdots \r_{(\td -3)}}  \pa_{\r_1} \cdots \pa^{\r_{(\td -3)}}.
\end{align}
Using Eq.(\ref{tenskaf}) and Eq.(\ref{tensk3}), we can write
\begin{align} \la{tensktotal2}
(K_F^{(\bd)})_{\m\n\a\b\w_1 \cdots \w_{(\bd -4)}} &= - \fr{\xi_{(d)}}{M_\g^{\bd -4}} \epsilon_{\m\n\r\l} \epsilon_{\a\b\s\t} \eta^{\t\l} u^\r u^\s X_{\w_1 \cdots \w_{(\bd -4)}} ; \nonumber\\
(K_{AF}^{(\td)})_\k^{\;\; \a_1 \cdots \a_{(\td -3)}} &= \fr{\xi_{(d')}}{M_\g^{\td -4}} u_\k Q^{\a_1 \cdots \a_{(\td-3)}} ,
\end{align}
where $Q$ and $X$ are symmetric tensors.

Through an rapid analysis, one can easily notices that the generalized terms proposed in Eq.(\ref{lagpargeral}) and Eq.(\ref{lagimpargeral}) can be rewritten as the coefficients defined above, with $Q$ and $X$ written in terms of the background fourvector $u^\l$. 

\bibliographystyle{apsrev4-1}
\bibliography{BIBLIOGRAFIA}

\begin{thebibliography}{32}%
\makeatletter
\providecommand \@ifxundefined [1]{%
 \@ifx{#1\undefined}
}%
\providecommand \@ifnum [1]{%
 \ifnum #1\expandafter \@firstoftwo
 \else \expandafter \@secondoftwo
 \fi
}%
\providecommand \@ifx [1]{%
 \ifx #1\expandafter \@firstoftwo
 \else \expandafter \@secondoftwo
 \fi
}%
\providecommand \natexlab [1]{#1}%
\providecommand \enquote  [1]{``#1''}%
\providecommand \bibnamefont  [1]{#1}%
\providecommand \bibfnamefont [1]{#1}%
\providecommand \citenamefont [1]{#1}%
\providecommand \href@noop [0]{\@secondoftwo}%
\providecommand \href [0]{\begingroup \@sanitize@url \@href}%
\providecommand \@href[1]{\@@startlink{#1}\@@href}%
\providecommand \@@href[1]{\endgroup#1\@@endlink}%
\providecommand \@sanitize@url [0]{\catcode `\\12\catcode `\$12\catcode
  `\&12\catcode `\#12\catcode `\^12\catcode `\_12\catcode `\%12\relax}%
\providecommand \@@startlink[1]{}%
\providecommand \@@endlink[0]{}%
\providecommand \url  [0]{\begingroup\@sanitize@url \@url }%
\providecommand \@url [1]{\endgroup\@href {#1}{\urlprefix }}%
\providecommand \urlprefix  [0]{URL }%
\providecommand \Eprint [0]{\href }%
\providecommand \doibase [0]{http://dx.doi.org/}%
\providecommand \selectlanguage [0]{\@gobble}%
\providecommand \bibinfo  [0]{\@secondoftwo}%
\providecommand \bibfield  [0]{\@secondoftwo}%
\providecommand \translation [1]{[#1]}%
\providecommand \BibitemOpen [0]{}%
\providecommand \bibitemStop [0]{}%
\providecommand \bibitemNoStop [0]{.\EOS\space}%
\providecommand \EOS [0]{\spacefactor3000\relax}%
\providecommand \BibitemShut  [1]{\csname bibitem#1\endcsname}%
\let\auto@bib@innerbib\@empty
\bibitem [{\citenamefont {Kosteleck{\`y}}(2004)}]{kostelecky2004}%
  \BibitemOpen
  \bibfield  {author} {\bibinfo {author} {\bibfnamefont {V.~A.}\ \bibnamefont
  {Kosteleck{\`y}}},\ }\href@noop {} {\bibfield  {journal} {\bibinfo  {journal}
  {Physical Review D}\ }\textbf {\bibinfo {volume} {69}},\ \bibinfo {pages}
  {105009} (\bibinfo {year} {2004})}\BibitemShut {NoStop}%
\bibitem [{\citenamefont {Liberati}\ and\ \citenamefont
  {Maccione}(2009)}]{liberati2009}%
  \BibitemOpen
  \bibfield  {author} {\bibinfo {author} {\bibfnamefont {S.}~\bibnamefont
  {Liberati}}\ and\ \bibinfo {author} {\bibfnamefont {L.}~\bibnamefont
  {Maccione}},\ }\href {\doibase 10.1146/annurev.nucl.010909.083640} {\bibfield
   {journal} {\bibinfo  {journal} {Ann. Rev. Nucl. Part. Sci.}\ }\textbf
  {\bibinfo {volume} {59}},\ \bibinfo {pages} {245} (\bibinfo {year} {2009})},\
  \Eprint {http://arxiv.org/abs/0906.0681} {arXiv:0906.0681 [astro-ph.HE]}
  \BibitemShut {NoStop}%
\bibitem [{\citenamefont {Collins}\ \emph {et~al.}(2004)\citenamefont
  {Collins}, \citenamefont {Perez}, \citenamefont {Sudarsky}, \citenamefont
  {Urrutia},\ and\ \citenamefont {Vucetich}}]{collins2004}%
  \BibitemOpen
  \bibfield  {author} {\bibinfo {author} {\bibfnamefont {J.}~\bibnamefont
  {Collins}}, \bibinfo {author} {\bibfnamefont {A.}~\bibnamefont {Perez}},
  \bibinfo {author} {\bibfnamefont {D.}~\bibnamefont {Sudarsky}}, \bibinfo
  {author} {\bibfnamefont {L.}~\bibnamefont {Urrutia}}, \ and\ \bibinfo
  {author} {\bibfnamefont {H.}~\bibnamefont {Vucetich}},\ }\href@noop {}
  {\bibfield  {journal} {\bibinfo  {journal} {Physical review letters}\
  }\textbf {\bibinfo {volume} {93}},\ \bibinfo {pages} {191301} (\bibinfo
  {year} {2004})}\BibitemShut {NoStop}%
\bibitem [{\citenamefont {Belich}\ \emph {et~al.}(2007)\citenamefont {Belich},
  \citenamefont {Costa-Soares}, \citenamefont {Santos},\ and\ \citenamefont
  {Orlando}}]{belich2007}%
  \BibitemOpen
  \bibfield  {author} {\bibinfo {author} {\bibfnamefont {H.}~\bibnamefont
  {Belich}}, \bibinfo {author} {\bibfnamefont {T.}~\bibnamefont
  {Costa-Soares}}, \bibinfo {author} {\bibfnamefont {M.}~\bibnamefont
  {Santos}}, \ and\ \bibinfo {author} {\bibfnamefont {M.}~\bibnamefont
  {Orlando}},\ }\href@noop {} {\bibfield  {journal} {\bibinfo  {journal}
  {Revista Brasileira de Ensino de F{\'\i}sica}\ }\textbf {\bibinfo {volume}
  {29}},\ \bibinfo {pages} {57} (\bibinfo {year} {2007})}\BibitemShut {NoStop}%
\bibitem [{\citenamefont {Ashtekar}(1986)}]{ashtekar1986}%
  \BibitemOpen
  \bibfield  {author} {\bibinfo {author} {\bibfnamefont {A.}~\bibnamefont
  {Ashtekar}},\ }\href {\doibase 10.1103/PhysRevLett.57.2244} {\bibfield
  {journal} {\bibinfo  {journal} {Phys. Rev. Lett.}\ }\textbf {\bibinfo
  {volume} {57}},\ \bibinfo {pages} {2244} (\bibinfo {year}
  {1986})}\BibitemShut {NoStop}%
\bibitem [{\citenamefont {Rovelli}(1998)}]{rovelli1997}%
  \BibitemOpen
  \bibfield  {author} {\bibinfo {author} {\bibfnamefont {C.}~\bibnamefont
  {Rovelli}},\ }\href {\doibase 10.12942/lrr-1998-1} {\bibfield  {journal}
  {\bibinfo  {journal} {Living Rev. Rel.}\ }\textbf {\bibinfo {volume} {1}},\
  \bibinfo {pages} {1} (\bibinfo {year} {1998})},\ \Eprint
  {http://arxiv.org/abs/gr-qc/9710008} {arXiv:gr-qc/9710008 [gr-qc]}
  \BibitemShut {NoStop}%
\bibitem [{\citenamefont {Ho{\v{r}}ava}(2009)}]{horava2009}%
  \BibitemOpen
  \bibfield  {author} {\bibinfo {author} {\bibfnamefont {P.}~\bibnamefont
  {Ho{\v{r}}ava}},\ }\href@noop {} {\bibfield  {journal} {\bibinfo  {journal}
  {Journal of High Energy Physics}\ }\textbf {\bibinfo {volume} {2009}},\
  \bibinfo {pages} {020} (\bibinfo {year} {2009})}\BibitemShut {NoStop}%
\bibitem [{\citenamefont {Kosteleck{\`y}}\ and\ \citenamefont
  {Samuel}(1989)}]{kostelecky1989}%
  \BibitemOpen
  \bibfield  {author} {\bibinfo {author} {\bibfnamefont {V.~A.}\ \bibnamefont
  {Kosteleck{\`y}}}\ and\ \bibinfo {author} {\bibfnamefont {S.}~\bibnamefont
  {Samuel}},\ }\href@noop {} {\bibfield  {journal} {\bibinfo  {journal}
  {Physical Review D}\ }\textbf {\bibinfo {volume} {39}},\ \bibinfo {pages}
  {683} (\bibinfo {year} {1989})}\BibitemShut {NoStop}%
\bibitem [{\citenamefont {Carroll}\ \emph {et~al.}(1990)\citenamefont
  {Carroll}, \citenamefont {Field},\ and\ \citenamefont
  {Jackiw}}]{carroll1990}%
  \BibitemOpen
  \bibfield  {author} {\bibinfo {author} {\bibfnamefont {S.~M.}\ \bibnamefont
  {Carroll}}, \bibinfo {author} {\bibfnamefont {G.~B.}\ \bibnamefont {Field}},
  \ and\ \bibinfo {author} {\bibfnamefont {R.}~\bibnamefont {Jackiw}},\
  }\href@noop {} {\bibfield  {journal} {\bibinfo  {journal} {Physical Review
  D}\ }\textbf {\bibinfo {volume} {41}},\ \bibinfo {pages} {1231} (\bibinfo
  {year} {1990})}\BibitemShut {NoStop}%
\bibitem [{\citenamefont {Colladay}\ and\ \citenamefont
  {Kosteleck{\`y}}(1997)}]{colladay1997}%
  \BibitemOpen
  \bibfield  {author} {\bibinfo {author} {\bibfnamefont {D.}~\bibnamefont
  {Colladay}}\ and\ \bibinfo {author} {\bibfnamefont {V.~A.}\ \bibnamefont
  {Kosteleck{\`y}}},\ }\href@noop {} {\bibfield  {journal} {\bibinfo  {journal}
  {Physical Review D}\ }\textbf {\bibinfo {volume} {55}},\ \bibinfo {pages}
  {6760} (\bibinfo {year} {1997})}\BibitemShut {NoStop}%
\bibitem [{\citenamefont {Colladay}\ and\ \citenamefont
  {Kosteleck{\`y}}(1998)}]{colladay1998}%
  \BibitemOpen
  \bibfield  {author} {\bibinfo {author} {\bibfnamefont {D.}~\bibnamefont
  {Colladay}}\ and\ \bibinfo {author} {\bibfnamefont {V.~A.}\ \bibnamefont
  {Kosteleck{\`y}}},\ }\href@noop {} {\bibfield  {journal} {\bibinfo  {journal}
  {Physical Review D}\ }\textbf {\bibinfo {volume} {58}},\ \bibinfo {pages}
  {116002} (\bibinfo {year} {1998})}\BibitemShut {NoStop}%
\bibitem [{\citenamefont {Gambini}\ and\ \citenamefont
  {Pullin}(1999)}]{gambini1998}%
  \BibitemOpen
  \bibfield  {author} {\bibinfo {author} {\bibfnamefont {R.}~\bibnamefont
  {Gambini}}\ and\ \bibinfo {author} {\bibfnamefont {J.}~\bibnamefont
  {Pullin}},\ }\href {\doibase 10.1103/PhysRevD.59.124021} {\bibfield
  {journal} {\bibinfo  {journal} {Phys. Rev.}\ }\textbf {\bibinfo {volume}
  {D59}},\ \bibinfo {pages} {124021} (\bibinfo {year} {1999})},\ \Eprint
  {http://arxiv.org/abs/gr-qc/9809038} {arXiv:gr-qc/9809038 [gr-qc]}
  \BibitemShut {NoStop}%
\bibitem [{\citenamefont {Carroll}\ \emph {et~al.}(2001)\citenamefont
  {Carroll}, \citenamefont {Harvey}, \citenamefont {Kostelecky}, \citenamefont
  {Lane},\ and\ \citenamefont {Okamoto}}]{carroll2001}%
  \BibitemOpen
  \bibfield  {author} {\bibinfo {author} {\bibfnamefont {S.~M.}\ \bibnamefont
  {Carroll}}, \bibinfo {author} {\bibfnamefont {J.~A.}\ \bibnamefont {Harvey}},
  \bibinfo {author} {\bibfnamefont {V.~A.}\ \bibnamefont {Kostelecky}},
  \bibinfo {author} {\bibfnamefont {C.~D.}\ \bibnamefont {Lane}}, \ and\
  \bibinfo {author} {\bibfnamefont {T.}~\bibnamefont {Okamoto}},\ }\href
  {\doibase 10.1103/PhysRevLett.87.141601} {\bibfield  {journal} {\bibinfo
  {journal} {Phys. Rev. Lett.}\ }\textbf {\bibinfo {volume} {87}},\ \bibinfo
  {pages} {141601} (\bibinfo {year} {2001})},\ \Eprint
  {http://arxiv.org/abs/hep-th/0105082} {arXiv:hep-th/0105082 [hep-th]}
  \BibitemShut {NoStop}%
\bibitem [{\citenamefont {Burgess}\ \emph {et~al.}(2002)\citenamefont
  {Burgess}, \citenamefont {Cline}, \citenamefont {Filotas}, \citenamefont
  {Matias},\ and\ \citenamefont {Moore}}]{burgess2002}%
  \BibitemOpen
  \bibfield  {author} {\bibinfo {author} {\bibfnamefont {C.~P.}\ \bibnamefont
  {Burgess}}, \bibinfo {author} {\bibfnamefont {J.~M.}\ \bibnamefont {Cline}},
  \bibinfo {author} {\bibfnamefont {E.}~\bibnamefont {Filotas}}, \bibinfo
  {author} {\bibfnamefont {J.}~\bibnamefont {Matias}}, \ and\ \bibinfo {author}
  {\bibfnamefont {G.~D.}\ \bibnamefont {Moore}},\ }\href {\doibase
  10.1088/1126-6708/2002/03/043} {\bibfield  {journal} {\bibinfo  {journal}
  {JHEP}\ }\textbf {\bibinfo {volume} {03}},\ \bibinfo {pages} {043} (\bibinfo
  {year} {2002})},\ \Eprint {http://arxiv.org/abs/hep-ph/0201082}
  {arXiv:hep-ph/0201082 [hep-ph]} \BibitemShut {NoStop}%
\bibitem [{\citenamefont {Jacobson}\ \emph {et~al.}(2002)\citenamefont
  {Jacobson}, \citenamefont {Liberati},\ and\ \citenamefont
  {Mattingly}}]{jacobson2001}%
  \BibitemOpen
  \bibfield  {author} {\bibinfo {author} {\bibfnamefont {T.}~\bibnamefont
  {Jacobson}}, \bibinfo {author} {\bibfnamefont {S.}~\bibnamefont {Liberati}},
  \ and\ \bibinfo {author} {\bibfnamefont {D.}~\bibnamefont {Mattingly}},\
  }\href {\doibase 10.1103/PhysRevD.66.081302} {\bibfield  {journal} {\bibinfo
  {journal} {Phys. Rev.}\ }\textbf {\bibinfo {volume} {D66}},\ \bibinfo {pages}
  {081302} (\bibinfo {year} {2002})},\ \Eprint
  {http://arxiv.org/abs/hep-ph/0112207} {arXiv:hep-ph/0112207 [hep-ph]}
  \BibitemShut {NoStop}%
\bibitem [{\citenamefont {Shao}\ and\ \citenamefont {Ma}(2010)}]{shao2010}%
  \BibitemOpen
  \bibfield  {author} {\bibinfo {author} {\bibfnamefont {L.}~\bibnamefont
  {Shao}}\ and\ \bibinfo {author} {\bibfnamefont {B.-Q.}\ \bibnamefont {Ma}},\
  }\href {\doibase 10.1142/S0217732310034572} {\bibfield  {journal} {\bibinfo
  {journal} {Mod. Phys. Lett.}\ }\textbf {\bibinfo {volume} {A25}},\ \bibinfo
  {pages} {3251} (\bibinfo {year} {2010})},\ \Eprint
  {http://arxiv.org/abs/1007.2269} {arXiv:1007.2269 [hep-ph]} \BibitemShut
  {NoStop}%
\bibitem [{\citenamefont {Amelino-Camelia}\ and\ \citenamefont
  {Piran}(2001)}]{amelino2000}%
  \BibitemOpen
  \bibfield  {author} {\bibinfo {author} {\bibfnamefont {G.}~\bibnamefont
  {Amelino-Camelia}}\ and\ \bibinfo {author} {\bibfnamefont {T.}~\bibnamefont
  {Piran}},\ }\href {\doibase 10.1103/PhysRevD.64.036005} {\bibfield  {journal}
  {\bibinfo  {journal} {Phys. Rev.}\ }\textbf {\bibinfo {volume} {D64}},\
  \bibinfo {pages} {036005} (\bibinfo {year} {2001})},\ \Eprint
  {http://arxiv.org/abs/astro-ph/0008107} {arXiv:astro-ph/0008107 [astro-ph]}
  \BibitemShut {NoStop}%
\bibitem [{\citenamefont {Castorina}\ \emph {et~al.}(2004)\citenamefont
  {Castorina}, \citenamefont {Iorio},\ and\ \citenamefont
  {Zappala}}]{castorina2004}%
  \BibitemOpen
  \bibfield  {author} {\bibinfo {author} {\bibfnamefont {P.}~\bibnamefont
  {Castorina}}, \bibinfo {author} {\bibfnamefont {A.}~\bibnamefont {Iorio}}, \
  and\ \bibinfo {author} {\bibfnamefont {D.}~\bibnamefont {Zappala}},\
  }\bibfield  {booktitle} {\emph {\bibinfo {booktitle} {{GZK and surroundings.
  Proceedings, Cosmic Ray International Seminars, CRIS 2004, Catania, Italy,
  May 31-June 4, 2004}}},\ }\href {\doibase 10.1016/j.nuclphysbps.2004.10.012}
  {\bibfield  {journal} {\bibinfo  {journal} {Nucl. Phys. Proc. Suppl.}\
  }\textbf {\bibinfo {volume} {136}},\ \bibinfo {pages} {333} (\bibinfo {year}
  {2004})},\ \bibinfo {note} {[,333(2004)]},\ \Eprint
  {http://arxiv.org/abs/hep-ph/0407363} {arXiv:hep-ph/0407363 [hep-ph]}
  \BibitemShut {NoStop}%
\bibitem [{\citenamefont {Shao}\ and\ \citenamefont {Ma}(2011)}]{shao2011}%
  \BibitemOpen
  \bibfield  {author} {\bibinfo {author} {\bibfnamefont {L.}~\bibnamefont
  {Shao}}\ and\ \bibinfo {author} {\bibfnamefont {B.-Q.}\ \bibnamefont {Ma}},\
  }\href {\doibase 10.1103/PhysRevD.83.127702} {\bibfield  {journal} {\bibinfo
  {journal} {Phys. Rev.}\ }\textbf {\bibinfo {volume} {D83}},\ \bibinfo {pages}
  {127702} (\bibinfo {year} {2011})},\ \Eprint {http://arxiv.org/abs/1104.4438}
  {arXiv:1104.4438 [astro-ph.HE]} \BibitemShut {NoStop}%
\bibitem [{\citenamefont {Couturier}\ \emph {et~al.}(2013)\citenamefont
  {Couturier}, \citenamefont {Vasileiou}, \citenamefont {Jacholkowska},
  \citenamefont {Piron}, \citenamefont {Bolmont}, \citenamefont {Granot},
  \citenamefont {Stecker}, \citenamefont {Cohen-Tanugi},\ and\ \citenamefont
  {Longo}}]{couturier2013}%
  \BibitemOpen
  \bibfield  {author} {\bibinfo {author} {\bibfnamefont {C.}~\bibnamefont
  {Couturier}}, \bibinfo {author} {\bibfnamefont {V.}~\bibnamefont
  {Vasileiou}}, \bibinfo {author} {\bibfnamefont {A.}~\bibnamefont
  {Jacholkowska}}, \bibinfo {author} {\bibfnamefont {F.}~\bibnamefont {Piron}},
  \bibinfo {author} {\bibfnamefont {J.}~\bibnamefont {Bolmont}}, \bibinfo
  {author} {\bibfnamefont {J.}~\bibnamefont {Granot}}, \bibinfo {author}
  {\bibfnamefont {F.}~\bibnamefont {Stecker}}, \bibinfo {author} {\bibfnamefont
  {J.}~\bibnamefont {Cohen-Tanugi}}, \ and\ \bibinfo {author} {\bibfnamefont
  {F.}~\bibnamefont {Longo}},\ }in\ \href
  {https://inspirehep.net/record/1251511/files/arXiv:1308.6403.pdf} {\emph
  {\bibinfo {booktitle} {{Proceedings, 33rd International Cosmic Ray Conference
  (ICRC2013): Rio de Janeiro, Brazil, July 2-9, 2013}}}}\ (\bibinfo {year}
  {2013})\ p.\ \bibinfo {pages} {0127},\ \Eprint
  {http://arxiv.org/abs/1308.6403} {arXiv:1308.6403 [astro-ph.HE]} \BibitemShut
  {NoStop}%
\bibitem [{\citenamefont {Kislat}\ and\ \citenamefont
  {Krawczynski}(2017)}]{kislat2017}%
  \BibitemOpen
  \bibfield  {author} {\bibinfo {author} {\bibfnamefont {F.}~\bibnamefont
  {Kislat}}\ and\ \bibinfo {author} {\bibfnamefont {H.}~\bibnamefont
  {Krawczynski}},\ }\href {\doibase 10.1103/PhysRevD.95.083013} {\bibfield
  {journal} {\bibinfo  {journal} {Phys. Rev.}\ }\textbf {\bibinfo {volume}
  {D95}},\ \bibinfo {pages} {083013} (\bibinfo {year} {2017})},\ \Eprint
  {http://arxiv.org/abs/1701.00437} {arXiv:1701.00437 [astro-ph.HE]}
  \BibitemShut {NoStop}%
\bibitem [{\citenamefont {Myers}\ and\ \citenamefont {Pospelov}(2003)}]{myers}%
  \BibitemOpen
  \bibfield  {author} {\bibinfo {author} {\bibfnamefont {R.~C.}\ \bibnamefont
  {Myers}}\ and\ \bibinfo {author} {\bibfnamefont {M.}~\bibnamefont
  {Pospelov}},\ }\href@noop {} {\bibfield  {journal} {\bibinfo  {journal}
  {Physical Review Letters}\ }\textbf {\bibinfo {volume} {90}},\ \bibinfo
  {pages} {211601} (\bibinfo {year} {2003})}\BibitemShut {NoStop}%
\bibitem [{\citenamefont {Reyes}(2010)}]{reyes2010}%
  \BibitemOpen
  \bibfield  {author} {\bibinfo {author} {\bibfnamefont {C.~M.}\ \bibnamefont
  {Reyes}},\ }\href {\doibase 10.1103/PhysRevD.82.125036} {\bibfield  {journal}
  {\bibinfo  {journal} {Phys. Rev.}\ }\textbf {\bibinfo {volume} {D82}},\
  \bibinfo {pages} {125036} (\bibinfo {year} {2010})},\ \Eprint
  {http://arxiv.org/abs/1011.2971} {arXiv:1011.2971 [hep-ph]} \BibitemShut
  {NoStop}%
\bibitem [{\citenamefont {Carroll}\ and\ \citenamefont
  {Tam}(2008)}]{carroll2008}%
  \BibitemOpen
  \bibfield  {author} {\bibinfo {author} {\bibfnamefont {S.~M.}\ \bibnamefont
  {Carroll}}\ and\ \bibinfo {author} {\bibfnamefont {H.}~\bibnamefont {Tam}},\
  }\href {\doibase 10.1103/PhysRevD.78.044047} {\bibfield  {journal} {\bibinfo
  {journal} {Phys. Rev.}\ }\textbf {\bibinfo {volume} {D78}},\ \bibinfo {pages}
  {044047} (\bibinfo {year} {2008})},\ \Eprint {http://arxiv.org/abs/0802.0521}
  {arXiv:0802.0521 [hep-ph]} \BibitemShut {NoStop}%
\bibitem [{\citenamefont {Kostelecky}\ and\ \citenamefont
  {Mewes}(2007)}]{kosteleckymewes2007}%
  \BibitemOpen
  \bibfield  {author} {\bibinfo {author} {\bibfnamefont {V.~A.}\ \bibnamefont
  {Kostelecky}}\ and\ \bibinfo {author} {\bibfnamefont {M.}~\bibnamefont
  {Mewes}},\ }\href {\doibase 10.1103/PhysRevLett.99.011601} {\bibfield
  {journal} {\bibinfo  {journal} {Phys. Rev. Lett.}\ }\textbf {\bibinfo
  {volume} {99}},\ \bibinfo {pages} {011601} (\bibinfo {year} {2007})},\
  \Eprint {http://arxiv.org/abs/astro-ph/0702379} {arXiv:astro-ph/0702379
  [ASTRO-PH]} \BibitemShut {NoStop}%
\bibitem [{\citenamefont {Kostelecky}\ and\ \citenamefont
  {Mewes}(2009)}]{kosteleckymewes2009}%
  \BibitemOpen
  \bibfield  {author} {\bibinfo {author} {\bibfnamefont {V.~A.}\ \bibnamefont
  {Kostelecky}}\ and\ \bibinfo {author} {\bibfnamefont {M.}~\bibnamefont
  {Mewes}},\ }\href {\doibase 10.1103/PhysRevD.80.015020} {\bibfield  {journal}
  {\bibinfo  {journal} {Phys. Rev.}\ }\textbf {\bibinfo {volume} {D80}},\
  \bibinfo {pages} {015020} (\bibinfo {year} {2009})},\ \Eprint
  {http://arxiv.org/abs/0905.0031} {arXiv:0905.0031 [hep-ph]} \BibitemShut
  {NoStop}%
\bibitem [{\citenamefont {Greiner}\ and\ \citenamefont
  {Reinhardt}(2013)}]{greiner}%
  \BibitemOpen
  \bibfield  {author} {\bibinfo {author} {\bibfnamefont {W.}~\bibnamefont
  {Greiner}}\ and\ \bibinfo {author} {\bibfnamefont {J.}~\bibnamefont
  {Reinhardt}},\ }\href@noop {} {\emph {\bibinfo {title} {Field
  quantization}}}\ (\bibinfo  {publisher} {Springer Science \& Business
  Media},\ \bibinfo {year} {2013})\BibitemShut {NoStop}%
\bibitem [{\citenamefont {Scatena}\ and\ \citenamefont
  {Turcati}(2014)}]{scatena2014}%
  \BibitemOpen
  \bibfield  {author} {\bibinfo {author} {\bibfnamefont {E.}~\bibnamefont
  {Scatena}}\ and\ \bibinfo {author} {\bibfnamefont {R.}~\bibnamefont
  {Turcati}},\ }\href {\doibase 10.1103/PhysRevD.90.127703} {\bibfield
  {journal} {\bibinfo  {journal} {Phys. Rev.}\ }\textbf {\bibinfo {volume}
  {D90}},\ \bibinfo {pages} {127703} (\bibinfo {year} {2014})},\ \Eprint
  {http://arxiv.org/abs/1411.4549} {arXiv:1411.4549 [hep-th]} \BibitemShut
  {NoStop}%
\bibitem [{\citenamefont {Abdo}\ \emph {et~al.}(2009)\citenamefont {Abdo} \emph
  {et~al.}}]{abdo2009}%
  \BibitemOpen
  \bibfield  {author} {\bibinfo {author} {\bibfnamefont {A.~A.}\ \bibnamefont
  {Abdo}} \emph {et~al.} (\bibinfo {collaboration} {Fermi GBM, Fermi-LAT}),\
  }\href {\doibase 10.1126/science.1169101} {\bibfield  {journal} {\bibinfo
  {journal} {Science}\ }\textbf {\bibinfo {volume} {323}},\ \bibinfo {pages}
  {1688} (\bibinfo {year} {2009})}\BibitemShut {NoStop}%
\bibitem [{\citenamefont {Hernaski}\ and\ \citenamefont
  {Belich}(2014)}]{hernaski2014}%
  \BibitemOpen
  \bibfield  {author} {\bibinfo {author} {\bibfnamefont {C.}~\bibnamefont
  {Hernaski}}\ and\ \bibinfo {author} {\bibfnamefont {H.}~\bibnamefont
  {Belich}},\ }\href {\doibase 10.1103/PhysRevD.89.104027} {\bibfield
  {journal} {\bibinfo  {journal} {Phys. Rev.}\ }\textbf {\bibinfo {volume}
  {D89}},\ \bibinfo {pages} {104027} (\bibinfo {year} {2014})},\ \Eprint
  {http://arxiv.org/abs/1409.5742} {arXiv:1409.5742 [hep-th]} \BibitemShut
  {NoStop}%
\bibitem [{\citenamefont {Casana}\ \emph
  {et~al.}(2010{\natexlab{a}})\citenamefont {Casana}, \citenamefont
  {Ferreira},\ and\ \citenamefont {Silva}}]{Casana2010a}%
  \BibitemOpen
  \bibfield  {author} {\bibinfo {author} {\bibfnamefont {R.}~\bibnamefont
  {Casana}}, \bibinfo {author} {\bibfnamefont {M.~M.}\ \bibnamefont {Ferreira},
  \bibfnamefont {Jr}}, \ and\ \bibinfo {author} {\bibfnamefont {M.~R.~O.}\
  \bibnamefont {Silva}},\ }\href {\doibase 10.1103/PhysRevD.81.105015}
  {\bibfield  {journal} {\bibinfo  {journal} {Phys. Rev.}\ }\textbf {\bibinfo
  {volume} {D81}},\ \bibinfo {pages} {105015} (\bibinfo {year}
  {2010}{\natexlab{a}})},\ \Eprint {http://arxiv.org/abs/0910.3709}
  {arXiv:0910.3709 [hep-th]} \BibitemShut {NoStop}%
\bibitem [{\citenamefont {Casana}\ \emph
  {et~al.}(2010{\natexlab{b}})\citenamefont {Casana}, \citenamefont {Ferreira},
  \citenamefont {Gomes},\ and\ \citenamefont {dos Santos}}]{Casana2010b}%
  \BibitemOpen
  \bibfield  {author} {\bibinfo {author} {\bibfnamefont {R.}~\bibnamefont
  {Casana}}, \bibinfo {author} {\bibfnamefont {M.~M.}\ \bibnamefont
  {Ferreira}}, \bibinfo {author} {\bibfnamefont {A.~R.}\ \bibnamefont {Gomes}},
  \ and\ \bibinfo {author} {\bibfnamefont {F.~E.~P.}\ \bibnamefont {dos
  Santos}},\ }\href {\doibase 10.1103/PhysRevD.82.125006} {\bibfield  {journal}
  {\bibinfo  {journal} {Phys. Rev.}\ }\textbf {\bibinfo {volume} {D82}},\
  \bibinfo {pages} {125006} (\bibinfo {year} {2010}{\natexlab{b}})},\ \Eprint
  {http://arxiv.org/abs/1010.2776} {arXiv:1010.2776 [hep-th]} \BibitemShut
  {NoStop}%
\end{thebibliography}%


\end{document}